\documentclass[prl,twocolumn,showpacs,preprintnumbers,amsmath,amssymb]{revtex4}

\usepackage[dvips]{graphics}
\usepackage{tikz}

\begin{document}

\title{The Quantum Totalitarian Property and Exact Symmetries}

\author{Chiara Marletto and Vlatko Vedral}
\affiliation{Clarendon Laboratory, University of Oxford, Parks Road, Oxford OX1 3PU, United Kingdom and\\Centre for Quantum Technologies, National University of Singapore, 3 Science Drive 2, Singapore 117543 and\\
Department of Physics, National University of Singapore, 2 Science Drive 3, Singapore 117542}

\date{\today}

\begin{abstract}
\noindent We discuss a point, which from time to time has been doubted in the literature: all symmetries, such as those induced by the energy and momentum conservation laws, hold in quantum physics not just ``on average", as is sometimes claimed, but {\sl exactly} in each ``branch" of the wavefunction, expressed in the basis where the conserved observable is sharp. We note that for conservation laws to hold exactly for quantum systems in this sense (not just on average), it is necessary to assume the so-called ``totalitarian property of quantum theory", namely that any system capable of measuring a quantum observable must itself be quantised. Hence, if conservation laws are to hold exactly, the idea of a `classical measuring apparatus' (i.e., not subject to the branching structure) is untenable. We also point out that any other principle having a well-defined formulation within classical physics, such as the Equivalence principle, is also to be extended to the quantum domain in exactly the same way, i.e., branch by branch. 
\end{abstract}

\pacs{03.67.Mn, 03.65.Ud}

\maketitle                           

In their highly educational book on quantum paradoxes \cite{AHRO}, Aharonov and Rohrlich present the following quantum effect which they claim has no classical analogue (whence the apparent paradox). A particle in a box is prepared in one of the allowed eigenstates (with say $n$ nodes). Then the box is suddenly expanded by a distance $\delta L =\lambda/2 (1-\epsilon)$ where $\epsilon$ is small (compared to the unity). The most probable new state will be the one which has $n+1$ nodes, whose wavelength is therefore shorter than the original wavelength (by the amount $\lambda \epsilon/(n+1)$).  Aharonov and Rohrlich conclude that, although the expected value of energy does not change when the box is suddenly expanded, the most likely new energy is actually higher than the original energy. This, they claim, cannot happen classically, since, classically, the particle cannot gain energy by being in an expanding box (as in the Joule-Thomson effect, \cite{KIT}). 

We would like to argue that this example does not present the whole picture of energy conservation in quantum theory: the apparent paradox is due to an incomplete model, which does not consider quantising the box itself.  There is no paradox after all, as we shall explain, provided one considers a completely quantum model. We will discuss and illustrate with examples the general point that conservation laws in quantum physics (in the Aharonov and Rohrlich's example, energy is conserved) still hold exactly in every branch of the wavefunction, where by 'branch' we mean each term in the linear expansion of the wavefunction in the basis where the conserved quantity is sharp. 

This is true also when measurements are involved, {\sl provided that the measuring apparatus is also quantised}, in which case a measurement is just the formation of entanglement between the system and the apparatus. (Hereinafter, by measuring apparatus we shall simply mean another quantum system that is designed to entangled itself to the system, in the basis defined by the observable to be measured). Consider an even simpler example, generalising the Aharonov and Rohrlich apparent paradox.  Suppose the electromagnetic (EM) field is in an equal superposition of the vacuum and ten photons state of light (this might be challenging to prepare, but it is in principle possible). Upon measuring the number of photons, we could with probability one half end up with ten photons. The energy of the field seems to have increased as a result! Initially it is equal on average to $5\hbar \omega$ and, if ten photons are detected,
it becomes $10\hbar\omega$. On average, of course, one obtains the vacuum state (assumed to have zero energy) half of the times, and so the {\sl mean energy} is the same before and after the measurement. So far, the resolution of the apparent paradox is exactly as in the above example of Aharonov and Rohrlich: energy is always conserved on average. 

But there is more to quantum energy conservation than just its average conservation - provided one considers a fully quantised measuring apparatus. Without any loss of generality, suppose that the measuring apparatus works based on the photoelectric effect, in such a way that each photons excites one electron. When the quantum measuring apparatus of the photon number is coupled to the field in that state, that measuring apparatus will change its energy during the measurement, compensating exactly the recorded mode energy for the field.  Then, the vacuum state, when coupled to the measuring apparatus, will not excite a single electron, while the ten photon state will excite ten electrons (which is what later becomes the signal we observe). 
The total state of the light mode plus the measuring apparatus is then:
\begin{equation}
|0\rangle |E\rangle +|0\rangle |E+10 e\rangle \;.
\end{equation}
The first ket represents the field, which after the measurement is in the vacuum. The second ket represents the state of the measuring apparatus, whose energy is initially assumed to be $E$, and in the second branch increases by $10e$, which denotes the energy of 10 excited electrons. Hence, the EM field is in the vacuum state in both branches, but the measuring apparatus has gained exactly the energy equivalent of absorbing the ten photons in the second branch. Therefore, in each branch, the energy is conserved, exactly. This, incidentally, corroborates the idea that a possible approach to quantisation
of classical systems is to take all classical possibilities, where conservation laws hold exactly, and allow for their superpositions (as in e.g. the path integral approach, \cite{FEY1}). 

Let us consider another example but this time involving momentum conservation. Imagine a photon going through a beamsplitter. If it reflects, it changes its momentum and therefore the beamsplitter has to recoil back in order to keep the overall momentum conserved. If the photon goes through, then the beamsplitter does not recoil. The total state of the photon and the beamsplitter is then
\begin{equation}
|r\rangle |\alpha-\delta p \rangle +|t\rangle |\alpha\rangle
\end{equation}
where $|r\rangle$ and $|t\rangle$ are the reflected and transmitted states of the photon and $|\alpha-\delta p \rangle$ and $|\alpha \rangle$ the corresponding states of the beamsplitter. The momentum is again conserved in each branch. The reason why we can claim that the photon after the beam-splitter is in a superposition $|r\rangle  +|t\rangle$ and is not entangled to the beamsplitter is because the beamsplitter is a well localised macroscopic object and is therefore not in a momentum eigenstate (in other words, 
$\langle \alpha-\delta p |\alpha\rangle \approx 1$ for the usual beamsplitter). This is a well-known result in the theory of reference frames, \cite{SPE1, SPE2}.

Abstracting from these examples, we would like to argue that the procedure to generalise all other principles prescribing conservation of given observables in classical physics is to consider them as prescribing a branch-by-branch exact conservation in quantum physics.
This version stipulates that observables are not just conserved on average (which could misleadingly imply that while in each run of the experiment the conservation law might be broken, and that it is only in the limit of a large number of trials that exact conservation is recovered). In other words, conservation laws in quantum theory hold at the level of q-numbered observables: the conservation of an observable Q of a given system $S$, represented by an operator $\hat Q_S$,  is expressed as requiring that all allowed unitaries $U_{SR}$ on the system $S$ and on the rest of the universe $R$ (here we assume that the rest of the universe acts as the ultimate measurement apparatus) satisfy the following constraint:

$$
[\hat Q_S+ \hat Q_R, U_{SR}]=0
$$

\noindent where $\hat Q_R$ is the operator representing the observable Q of the rest of the universe. 

For conservation laws to hold exactly, therefore, one has to quantise whatever interacts with the quantum system, including the measuring apparatus. This of course causes the latter to become entangled with the system $S$, in accordance with the so-called {\sl totalitarian property} of quantum theory, \cite{DEU}: anything that couples to a superposed quantum object has itself to become entangled to it, if the coupling is in the superposed basis. \footnote{This is to be distinguished from the similarly called totalitarian principle of quantum physics by Gell-Mann, which states that anything that is not prohibited is compulsory.} This was emphasised by Schr\"odinger in his famous cat thought experiment. It also features in all other fully quantum accounts of the measurement process, starting with von Neumann, continuing with Wigner and culminating with Everett, DeWitt and Deutsch, \cite{VON, WIG, EVE, DEW, DEU1}. 

Finally, this point is relevant for an ongoing a discussion, about how to express the Equivalence Principle in the quantum domain \cite{BEI, HARDY, BRU1, BRU2}. In that context, we have argued that other principles originally formulated within a classical theory, such as the Einstein equivalence principle, should be extended to quantum systems in exactly the way outlined above, \cite{MAVE1, MAVE2}. Namely, if we have a spatial superposition in which a massive particle exists at two locations, then this generates two different gravitational fields at a given distant point (assuming that the gravitational field is treated quantum-mechanically and in the first linear order of approximation, so that the same standards apply as in quantum electrodynamics). A test particle located at that distant point would then accelerate in both branches, towards  the massive particle's respective locations. The state of the initial mass, the field and the test mass would then be:
\begin{equation}
|r_1\rangle |g_1 \rangle |a_1 \rangle +|r_2\rangle |g_2\rangle |a_2 \rangle
\end{equation}
The equivalence principle, which says that the gravitational field is indistinguishable (locally) from acceleration, applies in each of the superposed spatial branches. This is to be expected since each branch represents a classical gravitational scenario, where the position observable is sharp with the respective values. It is of course possible that this view of the equivalence principle will be experimentally invalidated (we do not have any experimental evidence in this domain to guide us), however, our point is that there is no prima facie reason to think that the equivalence principle is in conflict with quantum physics (any more than energy conservation is). We mention in passing that this principles underlies our recently proposed experiment to witness quantum effects in the gravitational field, \cite{MAVE1, MAVE2, BOSE}.

Any consistent account of classical conservation laws and symmetries must be handled in quantum theory by incorporating all the relevant degrees of freedom so that the conservation and symmetry become exact, expressed at the level of q-numbers, or branch-by-branch.  This mandates to include in our models all the relevant entanglements due to the interaction between the system and the environment, including measurement apparatuses. What about effective classical measurement apparatuses - those that appear not to have the ability to interfere? As is well-known, the most natural view of the emergence of classicality in quantum physics is through decoherence. Decoherence is a process by which a quantum system becomes entangled with its environment to the degree that it loses the ability to interfere in the basis in which it is entangled; this is the basis of classical states since no superpositions are allowed, once decoherence has occurred. This, as we argued, is the only view of classicality compatible with exact conservation laws and other principles, such as the equivalence principle, \cite{MAVE1, MAVE3} and even locality \cite{MAVE4}. According to decoherence theory all conservation laws hold exactly providing that every relevant system is quantised and this includes the environment that causes decoherence. 

In conclusion, we do not see any need to postulate that conservation laws, symmetries and general classical principles hold in quantum physics only on average, so long as we apply quantum theory consistently with its totalitarian property. Relaxing the laws to hold only on average is sometimes proposed in the literature as a solution of apparent paradoxes, which are in fact due to not applying quantum physics consistently to the measuring apparatus. When considering a complete model, which applies quantisation consistently, such problems evaporate. Still, there are several outstanding issues. Our discussion after all is confined within non-relativistic quantum physics. In quantum field theory, which is the ultimate description we currently have of fundamental interactions in nature, it is not entirely clear how to state our quantum version of classical principles. If for example we are dealing with the scattering-matrix formalism, then the energy and momentum conservation laws are a direct consequence of the Lorentz invariance of the scattering amplitudes \cite{WEI}. This agrees with our concept of conservations holding exactly in every branch. It seems that the formulation of principles we discussed here is most compatible with the path integral approach to quantum field theory, since the latter is based on a coherent quantum sum over all classical field configurations. Given, however, that there are open issues with all known field quantisations procedures, we leave this thought as a seed for a future programme of research. 

\textit{Acknowledgments}: CM and VV thank the John Templeton Foundation and the Eutopia Foundation. VV's research is supported by the National Research Foundation and the 
Ministry of Education in Singapore and administered by Centre for Quantum Technologies, National University of Singapore. This publication was made possible through the support of the ID 61466 grant from the John Templeton Foundation, as part of the The Quantum Information Structure of Spacetime (QISS) Project (qiss.fr). The opinions expressed in this publication are those of the authors and do not necessarily reflect the views of the John Templeton Foundation.


\begin{thebibliography}{1}
\bibitem{AHRO} Y. Aharonov, D. Rohrlich, Quantum Paradoxes, Wiley-VCH (2005).
\bibitem{KIT} C. Kittel, H. Kroemer, {\sl Thermal Physics}, Wiley, (1969). 
\bibitem{FEY1} Feynman, R. P.; Hibbs, A. R. (1965). {\sl Quantum Mechanics and Path Integrals}. New York: McGraw-Hill. ISBN 978-0-07-020650-2.
\bibitem{SPE1} G. Gour and R.W. Spekkens, New J. Phys. 10:033023 (2008);
\bibitem{SPE2} S. D. Bartlett, T. Rudolph, and R.W. Spekkens, Rev. Mod. Phys. 79(2):555-609 (2007).
\bibitem{VON} J. von Neumann, {\sl Mathematical Foundations of Quantum Mechanics}, Princeton Univ. Press., (1996).
\bibitem{WIG} E. P. Wigner, {\sl Remarks on the mind-body question}, in: I. J. Good, {\sl The Scientist Speculates}, London, Heinemann, (1961).
\bibitem{EVE} H. Everett, {\sl On the foundations of quantum mechanics}. PhD thesis, Department of Physics, Princeton University, Princeton, NJ, USA, (1957). 
\bibitem{DEW} B. DeWitt, {\sl A Global approach to quantum field theory}, International Series of Monographs on Physics, 2014.
\bibitem{DEU1} D. Deutsch, Int. J. Theor. Phys. 24, 1-41 (1985). https://doi.org/10.1007/BF00670071 
\bibitem{BEI} C. Anastopoulos and B. L. Hu, Class. Quant. Grav. 35, 035011 (2018).
\bibitem{HARDY} L. Hardy, {\sl Implementation of the Quantum Equivalence Principle}, arXiv:1903.01289, (2019). 
\bibitem{BRU1} M. Zych, \v C. Brukner, Nat. Phys., {\bf 14}, 1027-1031, (2018)
\bibitem{BRU2} Rosi, G. et al., Nat. Commun. {\bf 8}, 15529 (2017); most recently F. Giacomini et al, Nat. Commun. {\bf 10}, 494 (2019).
\bibitem{MAVE1} C. Marletto, V. Vedral, {\sl Sagnac interferometer and the quantum nature of gravity}, https://arxiv.org/pdf/2001.02777.pdf
\bibitem{MAVE2} C. Marletto and V. Vedral, Phys. Rev. Lett. 119, 240402 (2017).
\bibitem{BOSE} S. Bose, et al. Phys. Rev. Lett. 119, 240401 (2017).
\bibitem{MAVE3} C. Marletto,  V. Vedral, {\sl On the testability of the equivalence principle as a gauge principle detecting the gravitational $t^3$ phase}, arXiv:2004.11616, in press.
\bibitem{MAVE4}  C. Marletto, V. Vedral, {\sl The Aharonov-Bohm phase is locally generated (like all other quantum phases)}, https://arxiv.org/pdf/1906.03440.pdf 
\bibitem{WEI} S. Weinberg, {\sl The Quantum Theory of Fields}, (Cambridge University Press 1995). 
\end{thebibliography}
\end{document}